\documentclass[12pt]{elsarticle}
\usepackage{refmerge}
\usepackage{epsfig}
\usepackage{lineno}
\def\Ecm {\ensuremath{\rm E_{\rm c.m.}}}
\def\BR {\ensuremath{\mathcal B}}
\def\epem {\ensuremath{e^+ e^-}}
\def\pipi {\ensuremath{\pi^+\pi^-}}
\def\mevc {\ensuremath{\rm MeV/c}}
\def\mevcc {\ensuremath{\rm MeV/c^2}}
\def\piz {\ensuremath{\pi^0}}
\begin{document}
\date{\today}

\title{\bf{ \boldmath
STUDY OF THE PROCESS $e^+e^-\to \pi^+\pi^-\pi^+\pi^-$
IN THE C.M. ENERGY RANGE 920--1060 MEV WITH THE CMD-3 DETECTOR
}}

\author[adr1,adr2]{R.R.Akhmetshin}
\author[adr1,adr2]{A.N.Amirkhanov}
\author[adr1,adr2]{A.V.Anisenkov}
\author[adr1,adr2]{V.M.Aulchenko}
\author[adr1]{V.Sh.Banzarov}
\author[adr1]{N.S.Bashtovoy}
\author[adr1,adr2]{D.E.Berkaev}
\author[adr1,adr2]{A.E.Bondar}
\author[adr1]{A.V.Bragin}
\author[adr1,adr2]{S.I.Eidelman}
\author[adr1,adr2]{D.A.Epifanov}
\author[adr1,adr2,adr3]{L.B.Epshteyn}
\author[adr1,adr2]{A.L.Erofeev}
\author[adr1,adr2]{G.V.Fedotovich}
\author[adr1,adr2]{S.E.Gayazov}
\author[adr1,adr2]{A.A.Grebenuk}
\author[adr1,adr2]{S.S.Gribanov}
\author[adr1,adr2,adr3]{D.N.Grigoriev}
\author[adr1,adr2]{F.V.Ignatov}
\author[adr1,adr2]{V.L.Ivanov}
\author[adr1]{S.V.Karpov}
\author[adr1]{A.S.Kasaev}
\author[adr1,adr2]{V.F.Kazanin}
\author[adr1,adr2]{I.A.Koop}
\author[adr1,adr2]{O.A.Kovalenko}
\author[adr1,adr2]{A.A.Korobov}
\author[adr1,adr3]{A.N.Kozyrev}
\author[adr1,adr2]{E.A.Kozyrev}
\author[adr1,adr2]{P.P.Krokovny}
\author[adr1,adr2]{A.E.Kuzmenko}
\author[adr1,adr2]{A.S.Kuzmin}
\author[adr1,adr2]{I.B.Logashenko}
\author[adr1]{A.P.Lysenko}
\author[adr1,adr2]{P.A.Lukin}
\author[adr1]{K.Yu.Mikhailov}
\author[adr1]{V.S.Okhapkin}
\author[adr1]{Yu.N.Pestov}
\author[adr1,adr2]{E.A.Perevedentsev}
\author[adr1,adr2]{A.S.Popov}
\author[adr1,adr2]{G.P.Razuvaev}
\author[adr1]{Yu.A.Rogovsky}
\author[adr1]{A.A.Ruban}
\author[adr1]{N.M.Ryskulov}
\author[adr1,adr2]{A.E.Ryzhenenkov}
\author[adr1,adr2]{V.E.Shebalin}
\author[adr1,adr2]{D.N.Shemyakin}
\author[adr1,adr2]{B.A.Shwartz}
\author[adr1,adr2]{D.B.Shwartz}
\author[adr1,adr4]{A.L.Sibidanov}
\author[adr1]{Yu.M.Shatunov}
\author[adr1,adr2]{E.P.Solodov\fnref{tnot}}
\author[adr1]{V.M.Titov}
\author[adr1,adr2]{A.A.Talyshev}
\author[adr1]{A.I.Vorobiov}
\author[adr1,adr2]{Yu.V.Yudin}


\address[adr1]{Budker Institute of Nuclear Physics, SB RAS, 
Novosibirsk, 630090, Russia}
\address[adr2]{Novosibirsk State University, Novosibirsk, 630090, Russia}
\address[adr3]{Novosibirsk State Technical University, 
Novosibirsk, 630092, Russia}
\address[adr4]{University of Victoria, Victoria, BC, Canada V8W 3P6}
\fntext[tnot]{Corresponding author: solodov@inp.nsk.su}


%
\vspace{0.7cm}
\begin{abstract}
\hspace*{\parindent}
A cross section of the process $e^+e^- \to \pipi\pipi$ has been measured
using 6798$\pm$93 signal events from a data sample corresponding
to an integrated luminosity of  9.8 pb$^{-1}$
collected with the CMD-3 detector 
in the center-of-mass energy range 920--1060 MeV. 
The measured cross section exhibits an interference pattern of the 
$\phi(1020)\to\pipi\pipi$ decay with a non-resonant process 
$e^+e^- \to \pipi\pipi$, from which we obtain the branching fraction
of the doubly suppressed decays (by G-parity and OZI rule):\\
$\BR(\phi\to\pipi\pipi) = (6.5\pm2.7\pm1.6)\times 10^{-6}$. 
\end{abstract}

\maketitle
\baselineskip=17pt
\section{ \boldmath Introduction}
\hspace*{\parindent}
Production of four charged pions in $e^+e^-$ annihilation 
has been studied 
with good statistics with the CMD-2~\cite{4pivepp2m1} and SND 
detectors~\cite{4pivepp2m2} as well as using initial-state radiation (ISR)
with BaBar~\cite{isr4pi} at which a low (about 3\%) systematic uncertainty 
was achieved on the $e^+e^-\to\pipi\pipi$ cross section in the wide 
center-of-mass energy (\Ecm) range. Earlier experiments  are discussed 
in Ref.~\cite{cmd2a1pi}.

However, a detailed study of this cross section in the vicinity of the 
$\phi(1020)$ resonance peak was performed by  the CMD-2 detector only 
(\Ecm = 984--1060 MeV)~\cite{phi4picmd2}, and 
$\BR(\phi\to\pipi\pipi) = (4.0^{+2.8}_{-2.2})\times 10^{-6}$ 
has been obtained. The $\phi(1020)$ decay to four charged pions is  
doubly suppressed by G-parity and the OZI-rule, and new measurements are 
interesting.  

In this paper we report an analysis of the data sample of
9.8 pb$^{-1}$ collected at the CMD-3 detector
in the \Ecm = 920--1060 MeV energy range. 
The data were collected in the energy scan of 22 c.m. energy
points performed  at  the VEPP-2000 collider~\cite{vepp} and used 
for a precision study of the process 
$\epem\to\phi\to K_S^0 K_L^0$~\cite{cmd3kskl} and obtaining the world best 
upper limit for the $\epem\to\eta'(958)$ process~\cite{cmd3etap}.  

The general-purpose detector CMD-3 has been described in 
detail elsewhere~\cite{sndcmd3}. Its tracking system consists of a 
cylindrical drift chamber (DC)~\cite{dc} and double-layer multiwire 
proportional Z-chamber, both also used for a trigger, and both inside 
a thin (0.2~X$_0$) superconducting solenoid with a field of 1.3~T.
The electromagnetic calorimeter (EMC) includes three systems.
The liquid xenon (LXe) barrel calorimeter with a thickness of 5.4~X$_0$ has
a fine electrode structure, providing 1--2 mm spatial resolution~\cite{lxe}, 
and shares the cryostat vacuum volume with the superconducting solenoid.     
The barrel CsI crystal calorimeter with a thickness of 8.1~X$_0$ is placed
outside  the LXe calorimeter,  and the endcap BGO calorimeter with a 
thickness of 13.4~X$_0$ is installed inside the solenoid~\cite{cal}.
The luminosity is measured using events of Bhabha scattering 
at large angles~\cite{lum} with about 1\% accuracy. 
The c.m. energy has been monitored by using the
Back-Scattering-Laser-Light system~\cite{laser} with about 0.06 MeV 
systematic uncertainty.
To obtain a detection efficiency, we have developed Monte Carlo (MC) 
simulation of our detector based on the GEANT4~\cite{geant4} package, 
in which the interaction of generated particles with the detector and its 
response are implemented.
MC simulation includes soft-photon radiation by the electron or positron, 
calculated according to Ref.~\cite{kur_fad}.
\section{Selection of $e^+e^-\to \pipi\pipi$ events}
\label{select}
\hspace*{\parindent}

Candidates for the process under study are required to have 
three or four tracks of charged particles in the DC with the following requirements:
the ionization losses of each track in the DC to be consistent 
with the pion hypothesis;
a  track momentum is larger than 40~\mevc; a minimum distance 
from a track to the beam axis in the transverse  plane is less than 0.25 cm; 
and a minimum distance from a track to the center of the interaction region 
along the beam axis Z  is less than 12 cm. Reconstructed momenta 
and angles of the tracks for three- and four-track candidates are used
for further selection.
  
A background in the studied energy region comes from the processes 
$\epem\to\pipi\piz$, $\epem\to K_S^0 K_L^0$, and $\epem\to K^+K^-$  
with extra tracks from decays or nuclear interaction of pions or kaons, 
as well as from a conversion of the photons from $\piz$-decay 
in the detector material.
Charged kaons are efficiently suppressed by the ionization losses in the DC. 
To suppress neutral kaons, we remove events with invariant mass 
of any two pion candidates within 20 MeV from the $K^0$ mass and having total 
momentum inside the 
20~\mevc~ window of the expected kaon momentum for the  $\epem\to K_S^0 K_L^0$ 
reaction. To reduce the background from the reaction $\epem\to\pipi\piz$,  
we require a missing mass for any two pion candidates to be greater than two 
pion masses.  After this requirement 
the remaining contribution from three pions is less than 0.3\% to number of 
four-track candidates. 
\begin{figure}[tbh]
\begin{center}
\vspace{-0.2cm}
\includegraphics[width=1.0\textwidth]{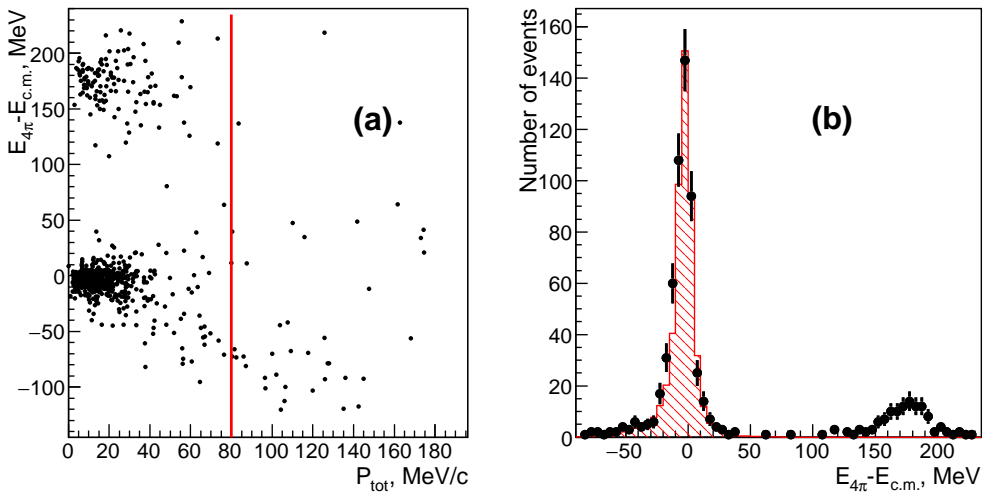}
\vspace{-0.3cm}
\caption
{
(a) Scatter plot of the difference  between the total energy  and
  c.m. energy (E$_{4\pi}$--\Ecm)
  versus total momentum for four-track events. The line shows 
the boundary of the applied selection; 
(b) Projection plot of (a) after selection. The histogram shows 
the MC-simulated distribution normalized to data.  
}
\label{4energy}
\end{center}
\end{figure}
\begin{figure}[tbh]
\begin{center}
\vspace{-0.2cm}
\includegraphics[width=1.0\textwidth]{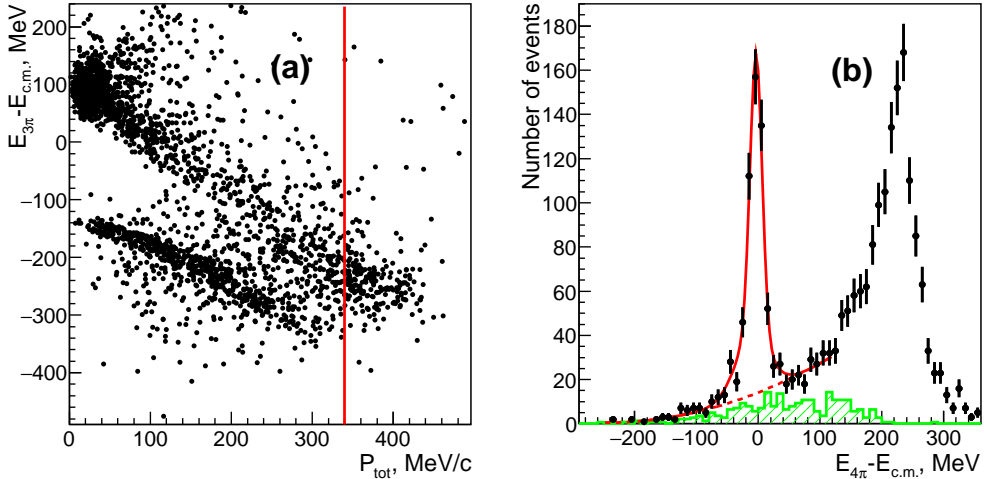}
\vspace{-0.3cm}
\caption
{
(a) Scatter plot of the difference  of the total energy 
and c.m. energy (E$_{3\pi}$--\Ecm)
  versus total momentum for three-track events; 
(b) Difference between the total energy of
  three tracks plus missing track energy  
  and c.m. energy (E$_{4\pi}$--\Ecm)(points). The line shows  a fit function 
used to obtain the number of signal events.
The shaded histogram shows 
an estimate of background events from the $\epem\to 3\pi$ process.
}
\label{3energy}
\end{center}
\end{figure}

For four- or three-track candidates we calculate the total energy and total
momentum assuming all tracks to be pions:
$$
\rm E_{3,\,4\pi} = \sum_{i=1}^{3,\,4}\sqrt{p_{i}^2+m_{\pi}^2}~,  ~~~~~P_{tot} =
\large |\sum_{i=1}^{3,\,4}\vec{ p}_{i}\large |.
$$

Figure~\ref{4energy}(a) shows a scatter plot of the difference between 
the total energy and c.m. energy E$_{4\pi} -\Ecm$ versus total
momentum for four-track candidates for \Ecm~ = 958 MeV. 
A clear signal of four-pion events is seen as a cluster of dots 
near zero.
Events with a radiative photon have non-zero total momentum and  total
energy, which is  always smaller than the nominal one.    
A momentum of any pion incorrectly reconstructed due to the interaction with
the detector material or DC resolution leads to momentum-energy
correlated ``tails'' in both directions.   

A cluster of dots is also observed shifted up from the four-pion signal. 
These events are from the process $\epem\to\epem\epem$ where electrons 
and positrons are produced due to two-photon processes as well as a conversion 
of radiative photons from the $\epem\to\epem\gamma$ reaction on the 
detector material, and a conversion of photons from the process 
$\epem\to\gamma\gamma$. 
Not all of these tracks can be identified as electron or positron in the EMC, but 
kinematically these events are well separated from the signal events, 
therefore we do not apply additional requirements.

We select events with total momentum less than 80 MeV/c and
show the difference E$_{4\pi}$--\Ecm~ in Fig.~\ref{4energy}(b). 
The experimental points are in good agreement with the
corresponding Monte Carlo simulated distribution shown by the histogram.
We require -100~$<$ E$_{4\pi} -\Ecm~ <$~50 MeV to determine the number of 
four-pion events, $N_{4 \rm tr}$. 
Four-track events have practically no background: we estimate it from MC 
simulation of the major background process $\epem\to\pipi\piz$
(a photon from the $\piz$ decay converts to an $e^+ e^-$ pair at the 
beam pipe),
and find a contribution of
less than 0.3\% at the peak of the $\phi$ resonance. We use this value as an 
estimate of the corresponding systematic uncertainty. 

To determine the number of four-pion events with one missing track, a sample
with three selected tracks is used. A track can be lost if it 
flies at  small polar angles outside the 
efficient DC region, decays in flight, due to incorrect
reconstruction, nuclear interactions or by overlapping with 
another track. 
Four-pion candidates in the three-track sample have energy
deficit correlated with the total (missing) momentum of three detected pions.
Figure~\ref{3energy}(a) shows a scatter plot of the difference 
E$_{3\pi}$--\Ecm~ between the total energy and c.m. energy versus total
momentum for three-track events. A band of signal events is clearly seen.
This sample has a large contribution from
the processes with electrons and positrons mentioned above as well as from  
the conversion of photons from the $\piz$ decays.
We apply an additional requirement on the maximum value of the total 
momentum of three tracks, assuming a four-pion final state. This requirement 
is shown by a line in Fig.~\ref{3energy}(a). 
 
Using total momentum of four-track candidates, we calculate the 
energy of a missing pion, and add it to the energy of three detected
pions: the difference of obtained energy and c.m. energy
is shown in Fig.~\ref{3energy}(b) by points.
The expected background contribution from three pions is shown by
the shaded histogram in Fig.~\ref{3energy}(b).

To obtain the number of four-pion events from the three-track sample,
we fit the distribution shown in Fig.~\ref{3energy}(b) with a sum of
the functions describing a signal peak and background.
The signal line shape is taken from the MC simulation 
of the four-pion process, and is well described by a sum of two 
Gaussian distributions.
The photon emission by initial electrons and positrons is taken into
account in the MC simulation, and gives a small asymmetry observed in the 
distributions of Fig.~\ref{4energy}(b) and Fig.~\ref{3energy}(b).
All parameters of the signal function are fixed except for the number of
events and the resolution of the narrowest Gaussian.
A second-order polynomial is used to describe the background distribution. 

Variation of the polynomial fit parameters for the experimental and 
MC-simulated background distributions, removing or applying background 
suppressing requirements lead to about 3\% uncertainty on 
the number of signal events.

We find 3690$\pm$61 four-track events and 3108$\pm$69 three-track events 
corresponding to the process $e^+e^-\to\pipi\pipi$ in the \Ecm~ = 920--1060 MeV
energy range. 
%
%
\begin{center}
\begin{figure}[tbh]
\vspace{-0.2cm}
\includegraphics[width=1.0\textwidth]{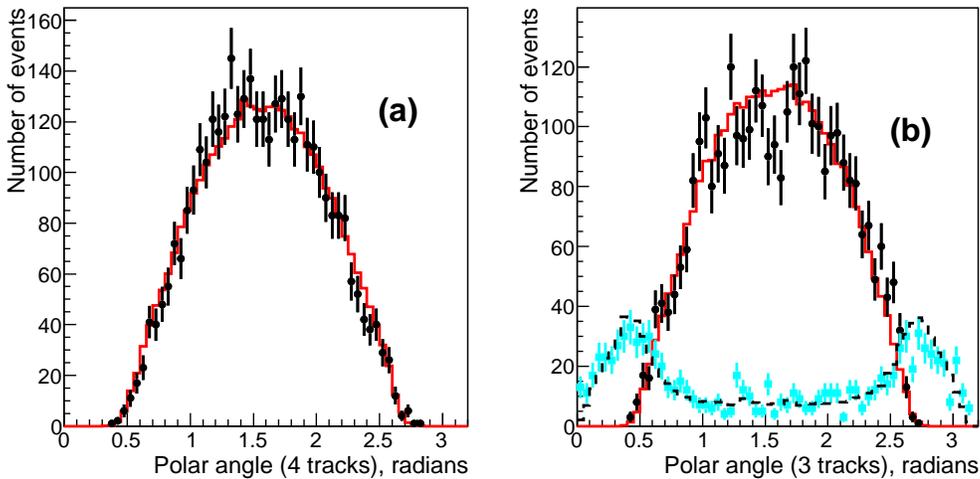}
\caption
{
(a) Polar angle distribution for four-pion events with four detected
  tracks for data (points) and MC simulation (histogram);
(b) Polar angle distribution for four-pion events with three detected
  tracks for data (circles) and MC simulation (solid histogram). 
The polar angle distribution for a missing track is shown by squares (data)
and the dashed histogram (MC simulation).
}
\label{angles}
\end{figure}
\end{center}
\section{Detection efficiency}
\label{efficiency}
\hspace*{\parindent}
In our experiment, the acceptance of the DC for the charged tracks is not 
100\%, and the detection efficiency depends on the dynamics of 
four-pion production.
The dynamics of the process $e^+e^-\to\pipi\pipi$ are relatively well 
studied in previous experiments~\cite{cmd2a1pi,isr4pi}, and the 
$a_1(1260)^{\pm}\pi^{\mp}$ final state has been shown to dominate. Our energy 
range is well below the nominal threshold of this reaction, and with our data sample 
we cannot observe any difference in any distribution for other final states, 
like $\rho(770) f_0(600)$ or phase-space production of four pions. 
   
Figure~\ref{angles}(a) presents the polar angle ($\theta_{\pi}$)
distribution for four-pion events with all detected tracks. 
The result of the MC simulation in the model with the 
$a_1(1260)\pi$ final state, presented by
the histogram well describes the observed distribution.
Figure~\ref{angles}(b) presents the polar angle distribution for
three detected tracks (circles for data, the solid
histogram for the MC simulation) after background subtraction.
The polar angle distribution for the missing track
is shown by squares (data) and the dashed histogram (MC).
With our DC acceptance we have about the same number of
four-pion events with one missing track compared to events
with all tracks detected.
\begin{center}
\begin{figure}[tbh]
\vspace{-0.2cm}
\includegraphics[width=1.0\textwidth]{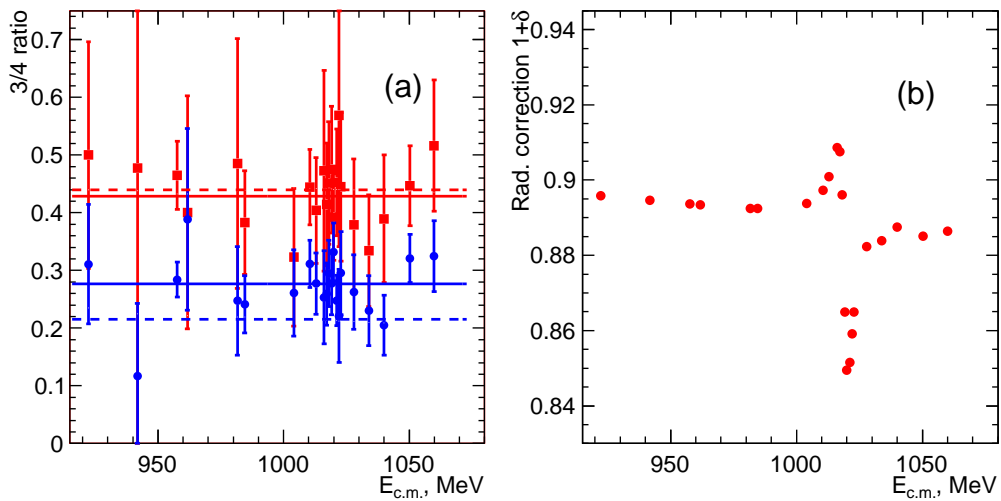}
\caption
{
(a) Ratio of numbers of  three- to four-detected-track events for data (points
  with errors) and MC simulation (the dashed line at 0.215) when all tracks 
are within the DC acceptance. The ratio of number of the events with one track outside 
the DC acceptance to number of the events with all tracks inside the DC acceptance 
(squares) and the corresponding MC-simulation value (the dashed line at 0.441).
(b) Radiative correction $1+\delta$ for the experimental energy points.
}
\label{ratio}
\end{figure}
\end{center}
Figure~\ref{ratio}(a) shows a ratio of the number of three-track events, 
$N_{3 \rm trDC}$, with a missing track inside a DC acceptance 
(0.7~$<\theta_{\pi}<$~2.44 radians) to the number of four-track events, 
$N_{4\rm tr}$, for data (circles) and 
MC-simulation (a dashed line at 0.215). This ratio for data exhibits very 
small variation with energy, but the average value of 0.274$\pm$0.012  differs 
from that for the MC-simulation. Based on these numbers we conclude 
that our MC-simulation overestimates a track reconstruction efficiency: 
0.951 vs 0.936$\pm$0.004 in data.
This difference does not depend on the primary generator model. We add events 
with a missing track inside the DC acceptance volume to a four-track sample, 
and this sum, $N_{4 \rm tr} +N_{3 \rm trDC} $,  corresponds to about 98\% of 
events with all four pions inside the DC acceptance: a probability to detect 
only one or two tracks from four is very low. A correction for the data-MC 
difference, $\epsilon1_{\rm corr}$, is about 1\%, and we assign a 
1\% systematic uncertainty to this value.
   
The number of remaining events with a missing track outside the DC 
acceptance, $N_{3 \rm tr}$, can be sensitive to the production dynamics and 
is used to validate MC-simulated efficiency. Figure~\ref{ratio}(a) 
shows a ratio of the number of three-track events with a missing track 
outside the DC acceptance to the sum of the numbers of four- and 
three-track events 
inside the DC acceptance for data (squares) and MC-simulation (the dashed line 
at 0.441). This ratio is also stable vs \Ecm, and the average value of 
0.429$\pm$0.022 is in agreement with that obtained from the MC-simulation.
The experimental value is also in agreement with the MC-simulation for the 
$\rho(770)f_0(600)$ final state (0.447) and with the phase-space model 
(0.435). Assuming single track reconstruction efficiency shown above, 
the data-MC correction for these three detected tracks, $\epsilon2_{\rm corr}$, 
is about $(5\pm3)$\%; the error is taken as a systematic uncertainty 
on this correction. 

We use the model with an $a_1(1260)\pi$ final state, and
calculate the detection efficiency from the MC-simulated events
as a ratio of all three- and four-track events to the total number of 
generated events. 
Note that if a sum of four- and three-track events 
is taken for the calculation, the efficiency increases to $\epsilon=$77.3 \%, 
independently of the c.m. energy, and 
the data--MC inconsistencies in the DC reconstruction efficiency 
and in the model-dependent angular distributions are significantly reduced.
\begin{center}
\begin{figure}[tbh]
\vspace{-0.2cm}
\includegraphics[width=1.0\textwidth]{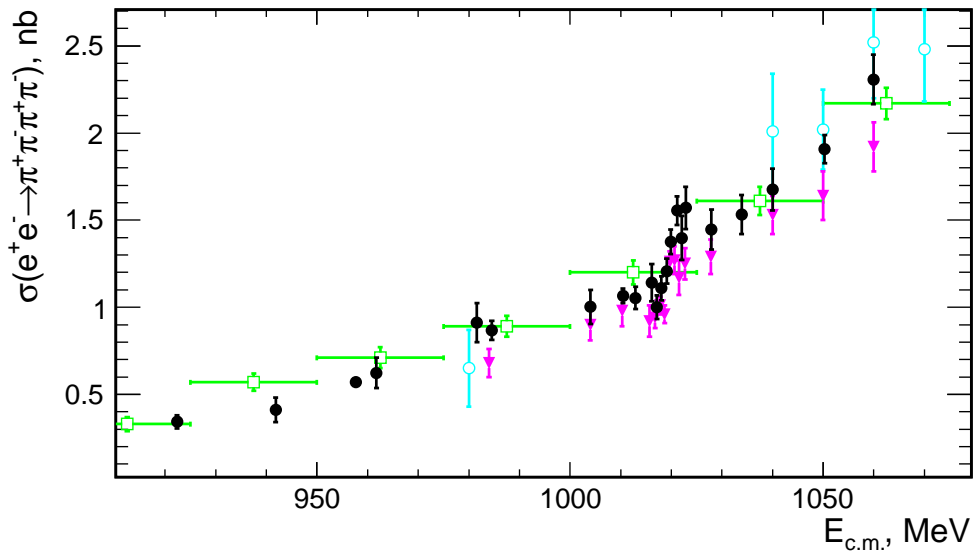}
\caption
{
The $e^+e^-\to\pipi\pipi$ cross section measured with the CMD-3
detector at VEPP-2000 (dots). The results of the BaBar
measurement~\cite{isr4pi} are shown by open squares, measurements by CMD-2 are shown by triangles~\cite{phi4picmd2} and by open circles~\cite{4pivepp2m1}.
}
\label{cross}
\end{figure}
\end{center}
\section{Cross Section Calculation}    
\hspace*{\parindent}
At each \Ecm~energy the cross section is calculated using four- and three-track events as 
$$
\sigma = \frac{(N_{4 \rm tr}+N_{3 \rm trDC})/(1-\epsilon1_{\rm corr})+N_{3 \rm tr}/(1-\epsilon2_{\rm corr})}{L\cdot\epsilon\cdot(1+\delta)},
$$ 
where $(1-\epsilon1_{\rm corr})$ is the data--MC correction to the number of 
events with four pions inside the DC  acceptance,  
$(1-\epsilon2_{\rm corr})$ is the the data--MC correction to the number of 
events with one pion out of the DC acceptance,
$L$ is the integrated luminosity at this energy, $\epsilon$ is the 
detection efficiency, 
and $(1+\delta)$ is the radiative correction calculated 
according to~\cite{kur_fad} and shown in Fig.~\ref{ratio} (b).
The energy dependence of the radiative correction reflects an interference 
pattern in the cross section. To calculate the correction, we use CMD-2
data~\cite{phi4picmd2} as a first approximation and then use our cross section 
data for the following iterations.  

Our result  is shown in Fig.~\ref{cross} by solid circles in comparison with the previous measurements.
The c.m. energy, integrated luminosity, number of four- and three-track events, radiative correction  and obtained cross section
for each energy are listed in Table~\ref{table}. 

The obtained cross section is in overall agreement with the results 
of the high-precision
measurement performed by the BaBar Collaboration~\cite{isr4pi}, shown
in Fig.~\ref{cross} by open squares, and reanalyzed data of 
CMD-2~\cite{4pivepp2m1} (open circles). Our cross section is about 
10\% higher than the CMD-2 measurement in the $\phi$-resonance 
region~\cite{phi4picmd2} (triangles), 
because these data were not reanalyzed, and the tracking efficiency 
was somehow overestimated.  

\begin{table}[tbh]
\caption{The c.m. energy, integrated luminosity, number of four- and three-track events, radiative correction, and $\epem\to\pipi\pipi$ cross section, measured with the CMD-3 detector. Only statistical errors are shown.
}
\label{table}
\begin{center}
\begin{tabular}{cccccc}
\hline
{\Ecm, MeV}
&{$L, \rm nb^{-1}$}
&{$N_{4\pi}$}
&{$N_{3\pi} +N_{3\pi DC}$}
&{$(1+\delta)$}
&{$\sigma$, nb}\\ 
\hline
922.35  & 414.1  & 49  & 47.0  $\pm$ 8.0  & 0.895 & 0.34 $\pm$ 0.04 \\ 
941.83  & 163.1  & 29  & 18.9  $\pm$ 5.5  & 0.893 & 0.41 $\pm$ 0.07 \\ 
957.68  & 2621   & 541 & 476.3 $\pm$ 26.3 & 0.892 & 0.57 $\pm$ 0.02 \\ 
961.75  & 128.1  & 27  & 25.5  $\pm$ 5.9  & 0.892 & 0.62 $\pm$ 0.09 \\ 
981.61  & 115.8  & 39  & 33.2  $\pm$ 6.4  & 0.891 & 0.91 $\pm$ 0.11 \\ 
984.54  & 468.1  & 162 & 116.3 $\pm$ 11.5 & 0.891 & 0.87 $\pm$ 0.05 \\ 
1004.07 & 195.4  & 80  & 53.3  $\pm$ 9.3  & 0.892 & 1.00 $\pm$ 0.10 \\ 
1010.47 & 936.1  & 352 & 314.6 $\pm$ 18.7 & 0.895 & 1.07 $\pm$ 0.04 \\ 
1012.96 & 485.4  & 194 & 153.5 $\pm$ 15.5 & 0.898 & 1.05 $\pm$ 0.06 \\ 
1016.15 & 192.1  & 82  & 69.2  $\pm$ 10.7 & 0.907 & 1.14 $\pm$ 0.11 \\ 
1017.16 & 479.0  & 185 & 144.7 $\pm$ 17.3 & 0.909 & 1.00 $\pm$ 0.07 \\ 
1018.05 & 478.3  & 191 & 167.8 $\pm$ 17.7 & 0.903 & 1.11 $\pm$ 0.07 \\ 
1019.12 & 477.9  & 202 & 178.4 $\pm$ 18.1 & 0.876 & 1.21 $\pm$ 0.07 \\ 
1019.90 & 570.2  & 269 & 229.7 $\pm$ 20.4 & 0.858 & 1.38 $\pm$ 0.07 \\ 
1021.16 & 475.4  & 265 & 215.3 $\pm$ 19.0 & 0.856 & 1.55 $\pm$ 0.08 \\ 
1022.08 & 201.6  & 97  & 88.8  $\pm$ 13.3 & 0.861 & 1.40 $\pm$ 0.13 \\ 
1022.85 & 195.3  & 106 & 92.4  $\pm$ 11.8 & 0.865 & 1.57 $\pm$ 0.12 \\ 
1027.96 & 195.8  & 108 & 80.0  $\pm$ 11.0 & 0.880 & 1.44 $\pm$ 0.11 \\ 
1033.91 & 175.5  & 110 & 70.6  $\pm$ 8.2  & 0.882 & 1.53 $\pm$ 0.11 \\ 
1040.03 & 195.9  & 133 & 89.5  $\pm$ 10.3 & 0.885 & 1.68 $\pm$ 0.12 \\ 
1050.31 & 499.6  & 323 & 294.5 $\pm$ 20.1 & 0.883 & 1.91 $\pm$ 0.08 \\ 
1059.95 & 198.9  & 147 & 148.6 $\pm$ 13.9 & 0.885 & 2.31 $\pm$ 0.14 \\ 
\hline
\end{tabular}
\end{center}
\end{table}

\section{Systematic errors}
\label{syst}
\hspace*{\parindent}
The following sources of systematic uncertainties are considered.

\begin{itemize}

\item {
Using three- and four-track events for the cross section calculation has 
reduced the overall model dependence uncertainty to about 1\%: the 
$a_1(1260)\pi$, $\rho(770)f_0(600)$ and phase-space models are tested. 
}  
\item{
Using responses of two independent triggers (neutral and charged) for our event sample, we found trigger efficiency close to unity with
a negligible contribution to the systematic error. 
}
\item{
The overall uncertainty on the determination of the integrated luminosity 
comes from the selection criteria of Bhabha events, radiative
corrections and calibrations of DC and calorimeters, and does not exceed 
1\%~\cite{lum}.
}
\item{
The admixture of the background events not subtracted from 
the four-track sample is estimated as a 0.3\% systematic uncertainty on 
the number of four-track events, with additional 1\% uncertainty 
from the data-MC efficiency correction.
}
\item{
The uncertainty on the background subtraction for three-track events is studied
by the variation of functions used for a background description in 
Fig.~\ref{3energy}(b) and is estimated as 3\% of the number of three-track 
events.
An additional 3\% uncertainty comes from the data-MC efficiency correction. 
}
\item{
A radiative correction uncertainty is estimated as about 1\%.
}
\end{itemize}

The above systematic uncertainties summed in quadrature and weighted with 
the number of three- and four-track events give an overall
systematic error of about 3.6\%. 
\begin{center}
\begin{figure}[tbh]
\vspace{-0.2cm}
\includegraphics[width=1.0\textwidth]{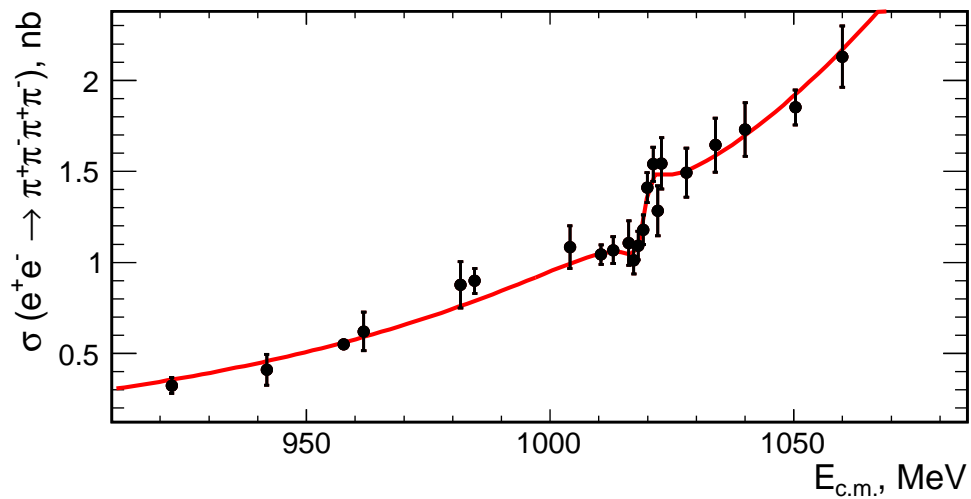}
\caption
{
The $e^+e^-\to\pipi\pipi$ cross section measured with the CMD-3
detector at VEPP-2000 (dots) with the fit function described in the text. 
}
\label{phifit}
\end{figure}
\end{center}
\section{Fit to $\phi\to\pipi\pipi$ decay rate}
\hspace*{\parindent}
The obtained cross section exhibits a clear interference pattern of the 
$\phi(1020)\to\pipi\pipi$ transition with the non-resonant cross section. 
We fit the experimental cross section with the function
$$
\sigma(\Ecm)=\sigma_0\cdot f(\Ecm)\cdot\left|1 - Z\cdot\frac{m_{\phi}\Gamma_{\phi}}{m^2_{\phi}-\Ecm^2-i\Ecm\Gamma_{\phi}}\right|^2,
$$
where $\sigma_0$ is a non-resonant cross section at the $\phi$ resonance mass 
$\Ecm = m_{\phi} = 1019.456$~\mevcc~ with $\Gamma_{\phi} = 4.24$ MeV 
width~\cite{pdg}, f(\Ecm) = $e^{A(\Ecm-m_{\phi})}$, $A$ 
is a slope parameter, describes the energy dependence of the non-resonant 
cross section, and Z is a complex amplitude of the 
$\phi(1020)\to\pipi\pipi$ transition. The fit yields

$\sigma_0 = 1.263 \pm 0.027~\rm nb, \\
\hspace*{\parindent} Re~Z = 0.146 \pm 0.030, \\
\hspace*{\parindent} Im~Z = -0.002 \pm 0.024. $ \\
The second solution gives the unphysical values for Z
and is ignored.
The $\phi\to\pipi\pipi$ decay rate is calculated as 

$\BR(\phi\to\pipi\pipi) = \sigma_0\cdot|\rm Z|^2/\sigma_{\phi} = (6.5\pm2.7\pm1.6)\times 10^{-6}$, \\
where $\sigma_{\phi} = 12\pi\BR(\phi\to\epem)/m_{\phi}^2 = 4172\pm 42$~nb is 
a peak cross section of $\phi(1020)$. The first error is statistical, while 
the second error is our estimate of the systematic uncertainty, based on the 
uncertainty on the cross section discussed in Sec.~\ref{syst}. 
The \Ecm~ energy spread (about 300 keV) contributes less than 4\% 
to the observed $\phi$ signal, and is negligibly small compared 
to other uncertainties.
Figure~\ref{phifit} shows our experimental points with the fit curve. 
The result is in overall agreement with the previous measurement 
$\BR(\phi\to\pipi\pipi) = (4.0^{+2.8}_{-2.2})\times 10^{-6}$~\cite{phi4picmd2}, 
and does not contradict to the value, assuming a single-photon reaction: 
 $\BR(\phi\to\gamma^*\to 4\pi) = 9\cdot\BR(\phi\to\epem)^2/\alpha^2 \cdot\sigma_0/\sigma_{\phi} = 4.8\times 10^{-6}$, which assumes Im Z=0.

\section*{ \boldmath Conclusion}
\hspace*{\parindent}
The total cross section of the process $e^+e^-\to\pipi\pipi$ 
has been measured using a data sample corresponding to an 
integrated luminosity of 9.8 pb$^{-1}$ 
collected by the CMD-3 detector at the VEPP-2000 $e^+e^-$ collider
in the 920--1060 MeV c.m. energy range. 
The three- and four-track events are used to estimate 
the model-dependent and other uncertainties on the cross section calculation, 
giving a 3.6\% overall systematic uncertainty.
The measured cross section is in overall agreement with 
previous experiments in the energy range studied, exhibits the 
interference pattern of the $\phi(1020)\to\pipi\pipi$ transition with 
the non-resonant cross section, and a new value 
$\BR(\phi\to\pipi\pipi) = (6.5\pm2.7\pm1.6)\times 10^{-6}$ has been obtained.

\subsection*{Acknowledgments}
\hspace*{\parindent}
We thank the VEPP-2000 personnel for the excellent machine operation.
Part of this work is supported by the Russian Foundation for Basic Research 
grants RFBR 15-02-05674-a and RFBR 16-02-00160-a.

\end{document}